\documentclass[aps,pra,showpacs,twocolumn]{revtex4}
\usepackage{graphicx}
\usepackage{epstopdf} 
\usepackage{amsmath}
\usepackage{amsfonts}

\begin{document}

\date{\today}
\author{J. Mumford}
\affiliation{Department of Physics and Astronomy, McMaster University, 1280 Main St.\ W., Hamilton, ON, L8S 4M1, Canada}
\author{D. H. J. O'Dell}
\affiliation{Department of Physics and Astronomy, McMaster University, 1280 Main St.\ W., Hamilton, ON, L8S 4M1, Canada}

\title{Critical exponents for an impurity in a bosonic Josephson junction: position measurement as a phase transition}

\begin{abstract}
We use fidelity susceptibility to calculate quantum
critical scaling exponents for a system consisting of $N$ identical
bosons interacting  with a single impurity atom in a double well
potential (bosonic Josephson junction).  Above a critical value of the boson-impurity interaction energy there is a spontaneous breaking of $\mathbb{Z}_2$ symmetry corresponding to a second order quantum phase transition from a balanced to an imbalanced number of particles in either the left or right hand well. 
We show that the exponents match those in the
Lipkin-Meshkov-Glick and Dicke models suggesting that the impurity model is in the same universality class.  The phase transition can be interpreted as a measurement of the position of the impurity by the bosons.
\end{abstract}

\pacs{03.75.Lm, 03.65.Ta, 67.85.Pq, 05.30.Rt}

\maketitle

\section{Introduction}
\label{sec:introduction}

\begin{figure}[t]
\begin{center}
\includegraphics[width=0.8\columnwidth]{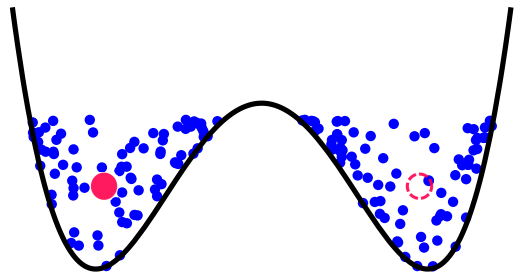}
\end{center}
\caption{(Color online) Schematic of the proposed set-up. A bosonic Josephson junction consists of $N$ identical bosons (represented by the small blue dots) which are able to tunnel between the two sides of a double well potential.  To this is added a single impurity atom (large red dot) which is also able to tunnel between the two wells.}
	\label{fig:doublewell}
\end{figure}

The fate of a single particle tunnelling in a many-body environment is a subject of fundamental interest not least because of its connection to the decoherence problem in quantum mechanics \cite{caldeira,leggett87}.
In this paper we study a related system consisting of a single impurity atom tunnelling between the wells of a double well potential in the presence of $N$ indistinguishable bosonic atoms as illustrated schematically in Figure \ref{fig:doublewell}.  The bosons are also trapped in the double well potential and thus form a bosonic Josephson junction in their own right.  This setup can be considered to be an elementary example of a Bose-Fermi mixture although, because the statistics of the impurity do not matter, in practice it can be a boson of the same species but in a different internal state. The prospects for realizing such a system in the laboratory are reasonably promising: a large number of experiments have studied ultracold bosons trapped in external double well potentials \cite{shin04,wang05,albiez05,gati06,schumm05,levy07,maussang10,baumgartner10,leblanc2011}, and others have realized the same effective system in a single trap but where two internal states of the atoms are coupled by microwave/radio frequency fields (internal Josephson effect)  \cite{gross10,zibold10}. Adding a well defined number of impurities is not easy but there has been some progress in this direction in optical lattices \cite{will2011,scelle13}.

A theoretical analysis of a bosonic Josephson junction with an impurity has been given by Rinck and Bruder \cite{rinck11}, 
who found that by applying a tilt to the double well a multi-particle tunnelling resonance could be induced towards a state where the impurity was expelled to the higher lying well. Subsequently, we undertook a study comparing the mean-field and many-body properties and described the appearance of a pitchfork bifurcation in the ground state of the mean-field theory above a certain critical value $W_{c}$ of the boson-impurity interaction strength \cite{mulansky11}. The mean-field bifurcation arises from the spontaneous localization of the impurity in one of the wells together with a localization of a majority of bosons in the opposite well (assuming repulsive interactions). In the fully quantum version $W_{c}$ marks the onset of a splitting of the wave function into two coherent pieces in Fock space (the space spanned by the Fock states $\vert \Delta M, \Delta N \rangle$ corresponding to the number differences $\Delta M= M_{R}-M_{L}$ and $\Delta N = N_{R}-N_{L}$ between the left and right wells for the impurity and bosons, respectively). As $W$  is increased further the Fock space splitting increases, and for large $N$ can develop into a fully blown Schr\"{o}dinger cat state which is a superposition of two macroscopically distinguishable number differences of bosons. This state is associated with a saturation of the entanglement entropy between the impurity and the bosons at $S=k_{B} \ln 2$. The formation of a Schr\"{o}dinger cat state in a macroscopic measurement device as a result of its coupling to a microscopic system is usually considered to be an essential element of quantum measurement \cite{vonneumann,zurek}. One may therefore take the view that the bosons in the present system act as a quantum measurement device or meter which indicates the position of the microscopic impurity atom. This meter can be tuned between being microscopic (small $N$) and macroscopic (large $N$). The formation and collapse of the Schr\"{o}dinger cat state corresponds here to a symmetry breaking phase transition \cite{damnjanovic,mayburov,allahverdyan,bargill,ivanov}.

In another study \cite{mumford14},  we argued that in many respects the impurity system behaves like the celebrated Dicke model \cite{dicke54,emary03,garraway} for $N$ two-level atoms
coupled to a single mode of the electromagnetic field whose Hamiltonian takes the form 
\begin{equation}\label{dicke}
\hat{H}_\mathrm{Dicke}=\hbar \omega \hat{a}^{\dag} \hat{a}+ \omega_{0} \hat{S}_z+\frac{2}{\sqrt{N}} \lambda\left(\hat{a}+\hat{a}^\dagger\right)\hat{S}_{x}. 
\end{equation}
Here $\hat{a}$ and $\hat{a}^{\dag}$ annihilate and create, respectively, a photon of energy $\hbar \omega$ in the electromagnetic field and $\hat{S}_{x}$ and $\hat{S}_{z}$ are collective spin operators that arise from treating the two-level atoms, whose levels are separated by energy $\hbar \omega_{0}$, as pseudospins. $\hat{S}_{z}$ measures half the difference between the number of atoms in the excited state and the ground state and it eigenvalues lie in the range $-N /2 \ldots N /2$. $\hat{S}_{x}=(\hat{S}_{+}+\hat{S}_{-})/2$ measures the coherence between the excited and ground states of the atoms and $\hat{a}+\hat{a}^\dagger$ is proportional to the position operator for the harmonic oscillator associated with the electromagnetic field. In a related pseudospin formulation the Hamiltonian for the bosonic Josephson junction plus impurity can be written (see Section \ref{sec:Model} for details)
\begin{equation}
\hat{H}=2 N J^{a} \hat{S}^{a}_{z}+2 J \hat{S}_{z}+2  W \hat{S}^{a}_{x}\hat{S}_{x}
\label{eq:spinH}
\end{equation}
 where the superscript `$a$' denotes the impurity: $J$ and $J^{a}$ are the bare hopping  frequencies between the two wells for the bosons and impurity, respectively, and $W$ parameterizes the boson-impurity coupling strength.
In this form the impurity model is reminiscent of the Mermin central-spin model where a
distinguishable central-spin is surrounded by $N$ spins on a lattice which interact with the central spin with an effectively infinite range interaction so that all pairwise interactions have the same magnitude
 \cite{breuer04,hassanieh06,garmon11}.  
In the impurity model $\hat{S}_{z}$ measures the coherence of the
bosons between the two wells, or, equivalently, half the difference in
the number of bosons in the antisymmetric and symmetric modes formed,
respectively, from the odd and even combinations of the modes
associated with each well. $\hat{S}_{x}$ measures half the number
difference between the two wells, or, equivalently, the coherence
between their symmetric and antisymmetric combinations. $S_{x}^{a}$
and $S_{z}^{a}$ are the corresponding quantities for the impurity. In
the thermodynamic limit where $N \rightarrow \infty$, the ground state
of  the Dicke model undergoes a second order phase transition due to a
spontaneous breaking of $\mathbb{Z}_{2}$  symmetry at the critical
coupling strength $\lambda_{c}=\sqrt{\omega \omega_{0}}/2$
\cite{hepp,wang}. This phase transition (PT) bears a very close resemblance to the bifurcation that occurs in the impurity model at \cite{mulansky11,mumford14}
\begin{equation}
W_{c}=\sqrt{J J^{a}}/2 .
\label{eq:crit}
\end{equation}
 In the Dicke case the ground state below the transition ($\lambda < \lambda_{c}$) is known as the normal state and is characterized by $\langle \hat{S}_{x} \rangle =0 $ and $\langle \hat{a} + \hat{a}^{\dag} \rangle=0$, whereas the ground state above the transition is known as the superradiant state because it corresponds to a spontaneous macroscopic excitation of the electromagnetic field with both   $\langle \hat{S}_{x} \rangle \neq 0 $ and $\langle \hat{a} + \hat{a}^{\dag} \rangle \neq 0$. Analogous ground states occur for the impurity model: when $W< W_{c}$ both the boson and the impurity probability distributions are symmetric $\langle \hat{S}_{x} \rangle =0$ and  $\langle \hat{S}^{a}_{x} \rangle =0$ and both expectation values acquire finite values in the symmetry broken state occurring when $W> W_{c}$. Furthermore, the dependence of the ground state energy upon the scaled parameters $W/W_{c}$ and $\lambda/\lambda_{c}$ is identical in the two models in the immediate vicinity of the transition \cite{mumford14}. It is also notable that the mean-field dynamics is in both cases regular below the transition and chaotic above it \cite{emary03,mumford14}. In this paper we shall further investigate the bifurcation in the impurity model by calculating the critical exponents in order to establish whether it is indeed a second order phase transition in the same universality class as that in the Dicke model.

Although both the Dicke and Impurity models share many common features there is one glaring difference: the Dicke model couples $N$ spin-1/2 particles to a harmonic oscillator whereas the impurity model couples $N$ spin-1/2 particles to one other spin. In essence the impurity model truncates the harmonic oscillator Hilbert space to just two states, the ground state and the first excited state. The spin-1/2 representing the impurity can never become macroscopically excited like the simple harmonic oscillator can. It is therefore quite remarkable that the impurity model behaves like the Dicke model, but the critical exponents we calculate here show that very close to the transition a two-state Hilbert space for the harmonic oscillator in the Dicke model suffices to describe its critical properties.

In order to investigate the critical behavior and obtain the critical scaling exponents
we shall calculate the fidelity susceptibility of the ground state. Over the past decade the concept of fidelity, which originated in quantum information theory \cite{nielsen00}, has gained wide use in analyzing
critical behavior and classifying the universality of systems.  It is
most commonly used to quantify changes in the ground state of a system
over a PT.  This is done by calculating the product
between the ground state with itself at different points in parameter
space 
\begin{equation}
F(W,\delta W) = \vert \langle \psi_0(W)\vert \psi_0(W + \delta W)
\rangle \vert
\label{eq:fidelity}
\end{equation}
where $W$ is the tunable parameter that drives the PT and $\psi_0$ is
the ground state.  It is expected that $F(W,\delta W) $ will tend to
unity away from the critical region and reach a minimum when $W =
W_c - \delta W/2$ where the scalar product will be between the ground state below
and above the critical point.  One of the first phase transitions to be studied using the fidelity was the one-dimensional (1D) XY model where it was shown to decrease to a minimum near the critical point \cite{zanardi06}.
Furthermore, excited state
fidelity has been used to characterize quantum phase transitions (QPT)  where the ground state
fidelity has failed \cite{chen07}.  Since the fidelity is a quantity
depending only on the geometry of the Hilbert space and requires no
knowledge of the order parameter it is useful in cases where the
order parameter of a system is not obvious and has been
studied in a variety of systems \cite{buonsante07,quan06,gu08}.
That being said, a more sensitive and natural
quantity to study where no \textit{a priori} knowledge of the system
is needed, is the fidelity susceptibility (FS) \cite{you07,yang07}.
The FS measures the response of the fidelity to infinitesimal changes
in the driving parameter of the system.  It is
closely related to the second derivative of the ground state energy
with respect to the driving parameter, $\frac{\partial^2 E_0}{\partial W^2}$, so
the FS is also similar to the magnetic susceptibility or specific heat
when the 
driving parameters are the magnetic field and temperature, respectively.
This means the FS can be used to study the critical behaviour of a
system through calculations of scaling exponents.

In this paper we add to work done by others \cite{liu12, sowinski13,buonsante12}
regarding the scaling and criticality of bosons in a double well potential.  We follow standard steps \cite{ning08,kwok08}
to show that the FS can be used to calculate scaling exponents for a
general system.  We then use the FS to focus on the critical
behaviour of the two-site boson-impurity Hubbard model.  The paper is organized in the following way: In Sec \ref{sec:Model} we go into more detail about our model for the physical system under study.  In Sec \ref{sec:fs} we show how critical scaling exponents can be extracted from the FS.  In Sec \ref{sec:nr}
we apply the methods of Sec \ref{sec:fs} to our system as well as extrapolate data
to find numerical values for $W_c$.  In Sec \ref{sec:ar} we find the
FS critical exponents analytically and in Sec \ref{sec:dc} we give a summary and outlook for further work.  Some of the details of the analytic calculations have been placed in an appendix.

\section{Model}
\label{sec:Model}

We model the bosonic Josephson junction plus impurity system using the two-site Bose Hubbard Hamiltonian \cite{rinck11,mulansky11}
\begin{equation}
\hat{H} = - N J^a \hat{A} -J \hat{B} + \frac{W}{2} \Delta \hat{N}
\Delta \hat{M} \, . 
\label{eq:ham}
\end{equation}  
Here, $\Delta \hat{N} \equiv \hat{b}^{\dagger}_R
\hat{b}_R - \hat{b}^{\dagger}_L \hat{b}_L$ is the number difference
operator between the two wells for the bosons and $\hat{B} \equiv
\hat{b}^{\dagger}_L \hat{b}_R + \hat{b}^{\dagger}_R \hat{b}_L$ is
the boson hopping operator. $\Delta \hat{M} \equiv \hat{a}^{\dagger}_R
\hat{a}_R - \hat{a}^{\dagger}_L \hat{a}_L$ and $\hat{A} \equiv
\hat{a}^{\dagger}_L \hat{a}_R + \hat{a}^{\dagger}_R \hat{a}_L$ are
the equivalent operators for the impurity.  The $L$ and $R$
subscripts denote the left and right modes and the creation/annihilation
operators follow the usual bosonic commutation relations, i.e.\
$[\hat{b}_\alpha,\hat{b}_\alpha^\dagger]=[\hat{a}_\alpha,\hat{a}_\alpha^\dagger]=1$
with $\alpha=L,\,R$ and all other combinations of the boson and
impurity operators are zero.    The scaling by $N$ in the first term in Eq.\ (\ref{eq:ham}) is
applied so that every term is $\mathcal{O}(N)$ and  therefore 
$W_c$ takes a finite value in the thermodynamic limit.  The pseudospin  formulation of the Hamiltonian given in Eq.\ (\ref{eq:spinH}) is obtained from Eq.\ (\ref{eq:ham}) by introducing the symmetric and antisymmetric combinations of the $L$ and $R$ modes: 
$\hat{b}_L \equiv \frac{1}{\sqrt{2}} \left ( \hat{b}_{S} + \hat{b}_{AS} \right )$ and
$\hat{b}_R \equiv \frac{1}{\sqrt{2}} \left ( \hat{b}_{S} - \hat{b}_{AS} \right )$, and then
applying Schwinger's oscillator model for angular momentum \cite{sakurai} $\hat{S}_z \equiv ( \hat{b}^{\dagger}_{AS} \hat{b}_{AS} -
  \hat{b}^{\dagger}_{S} \hat{b}_{S} )/2 =-\hat{B}/2$ and
$\hat{S}_x \equiv  ( \hat{b}^{\dagger}_{AS} \hat{b}_{S} +
  \hat{b}^{\dagger}_{S} \hat{b}_{AS} )/2 =-\Delta \hat{N}/2$. An analogous set of transformations apply to the impurity.

We do not include direct boson-boson intra-well (or inter-well) interactions in our calculations
and assume they can be removed (or the boson-impurity interaction enhanced) by a Feshbach resonance if necessary.
We do this both to highlight the effect of the impurity and also because it turns out not to change the results in a qualitative way. Indeed, the nonlinearity due to the boson-boson interactions can lead to very similar results as those resulting from the boson-impurity interaction (the impurity can be viewed as mediating an effective interaction between the bosons). In the case of repulsive boson-boson interactions, a purely bosonic system has no PT in the ground state but does experience 
a symmetry breaking bifurcation in the excited states known as macroscopic self-trapping \cite{milburn97,smerzi97} which has been seen in experiments \cite{albiez05}. If, on the other hand, the boson-boson interactions are attractive then there is  a $\mathbb{Z}_{2}$ symmetry breaking PT in the ground state above a critical interaction strength where the bosons clump together in a single well. This PT has been studied by Buonsante \textit{et al} \cite{buonsante12} and we shall find that the PT in our system falls in the same universality class.

In previous work we found through stability analysis around the mean-field stationary
points \cite{mumford14} that a pitchfork bifurcation of $\Delta N$ occurs at a critical value of the
boson-impurity interaction $W_{c}$ given above in Eq.\ (\ref{eq:crit}). For $W < W_c$, $\Delta N = 0$ and the bosons occupy each well equally. Above $W_c$ it becomes energetically favourable for the bosons to
build-up in one well and the impurity to be localized in the opposite
well.  This transition corresponds to the breaking of the
$\mathbb{Z}_2$ symmetry characterized by 
\begin{equation}
\left ( \Delta \hat{M}, \Delta \hat{N}, \hat{A}, \hat{B} \right ) \rightarrow \left ( -\Delta \hat{M},- \Delta \hat{N}, \hat{A}, \hat{B} \right ) \, .
\end{equation}
We will consider $W$ as the driving parameter and will analyze the system's
response to infinitesimal changes in it through the FS. 

\section{Fidelity susceptibility}
\label{sec:fs}

\begin{figure}[t]
\begin{center}
\includegraphics[width=0.8\columnwidth]{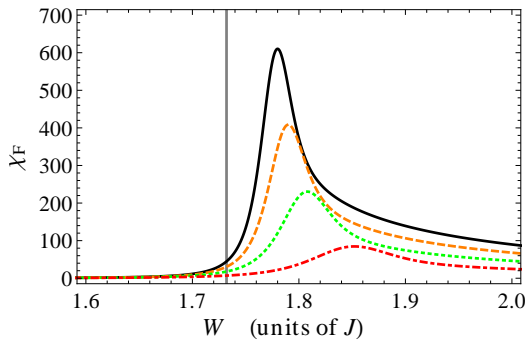}
\end{center}
\caption{(Color online) Fidelity susceptibility as a function of $W$ for
  different system sizes: $N= 200$ (red dot-dashed), $N=400$ (green dotted), $N=600$
(orange dashed), and $N = 800$ (black solid).  Here $J^a =
0.75 \, J$, so $W_c = \sqrt{3} \, J$ which is shown by the 
vertical line. It is clear that $\chi_{F}$ is not symmetric about the transition and hence the need for two indices $\pm \alpha$ as indicated in Eq.\ (\ref{eq:max}).}
	\label{fig:FS2}
\end{figure} 

As mentioned in the introduction, a more sensitive quantity than the
fidelity is the FS which we shall denote by $\chi_{\mathrm{F}}$. The two are related
through the Taylor expansion of Eq.\ (\ref{eq:fidelity}) to
second order
\begin{equation}
\textit{F}(W, \delta W) \approx 1 -
\frac{\chi_{\mathrm{F}}(W)}{2} (\delta W)^2+ ... \, .
\label{eq:TaylorF}
\end{equation}
It can be viewed as the system's response to an
infinitesimal change in the driving parameter. 
Equation (\ref{eq:ham}) has the general form
\begin{equation}
\hat{H} = \hat{H}_0 + W \hat{H}_I
\end{equation}
where $H_I$ is considered to be the driving term of the system.
From perturbation theory \cite{yang07} the FS is 
\begin{equation}
\chi_{\mathrm{F}}(W) = \sum_{n \neq 0} \frac{ \vert \langle \psi_n(W)
  \vert \hat{H}_I \vert \psi_0(W) \rangle \vert^2}{(E_n - E_0)^2} \, ,
\label{eq:FS}
\end{equation}
where $\psi_n(W)$ and $E_n$ are the $n$th eigenstate and
eigenenergy of the entire Hamiltonian, respectively. It is expected that for finite $N$ the FS scales as \cite{ning08,kwok08} 
\begin{equation}
\frac{\chi_{\mathrm{F}}}{N^d} \propto 1/ \vert W - W_{\mathrm{max}}
\vert^{\alpha_\pm}
\label{eq:max} 
\end{equation}
where $\alpha_\pm$ is the scaling
exponent above and below the quantum critical point (QCP),
respectively, $W_{\mathrm{max}}$ is the value of $W$ at which
$\chi_{\mathrm{F}}$ is at a maximum, and
$\chi_{\mathrm{F}} / N^d$ is an intensive quantity.  When $W =
W_{\mathrm{max}}$ $\chi_{\mathrm{F}}$ will be limited by the size of
the system, so we have
\begin{equation}
\chi_{\mathrm{Fmax}} \propto N^\mu \, .
\label{eq:mu}
\end{equation}
This quantity will diverge in the thermodynamic limit as
$W_{\mathrm{max}} \rightarrow W_c$. In fact, when Eq.\ (\ref{eq:ham})
is divided by $N$ so that each term is $\mathcal{O}(1)$ rather than $\mathcal{O}(N)$, then the exponent $\mu$ also
gives the scaling of the energy gap between the ground and first
excited states \cite{jacobson09, venuti07} as we have verified \cite{note1}. Figure \ref{fig:FS2} illustrates how $\chi_{\mathrm{Fmax}}$, which is given by the peak of each curve, depends on $N$.   In order to capture the behavior of both
Eq.\ (\ref{eq:max}) and Eq.\ (\ref{eq:mu}) we use the following form \cite{ning08}
\begin{equation}
\frac{\chi_{\mathrm{F}}}{N^d} = \frac{c}{N^{-\mu + d} + g(W) \vert W -
  W_{\mathrm{max}}\vert^\alpha} \, ,
\label{eq:universalFS1}
\end{equation}
where $c$ is a constant and $g(W)$ is a nonzero function of $W$, both
being intensive quantities.  Since we are dealing with the susceptibility of the ground state
wave function in the Fock basis, $N$ plays the role of the system
size.  With this in mind we can use the finite size scaling hypothesis
\cite{privman84} giving
\begin{equation}
f = N^{-1} Y \left [ N^a (W - W_{\mathrm{max}}) \right ]
\label{eq:fenergy}
\end{equation}
where $f$ is the free energy density and $Y$ is some
function.  We expect Eq.\ (\ref{eq:fenergy}) to vanish as $W
\rightarrow W_{\mathrm{max}}$ and at the same time the domain of the
correlations to diverge.  In this limit it is natural to expect \cite{pathria11}
\begin{equation}
f \sim \xi^{-1} \sim (W - W_{\mathrm{max}})^\nu
\label{eq:sim}
\end{equation}
where $\xi$ is the correlation length (in Fock space) and $\nu$ is the correlation length critical
exponent.  Combining Eqns.\ (\ref{eq:fenergy}) and (\ref{eq:sim})
gives the relation $a = 1/\nu$.  Using the fact that in
general the susceptibility due to $W$ is $\chi = - \frac{\partial^2
  f}{\partial W^2}$ we can show the reduced FS is a universal function
of $N$ and the driving parameter
\begin{equation}
\frac{\chi_{\mathrm{Fmax}} - \chi_{\mathrm{F}}}{\chi_{\mathrm{F}}} = X[N^{1/\nu}(W - W_{\mathrm{max}})]
\label{eq:universalFS2}
\end{equation}
where $X$ is some function.  Finally, combining this equation with Eq.\
(\ref{eq:universalFS1}) gives us the important scaling relation
\begin{equation}
\alpha = \nu (\mu - d) 
\label{eq:scalerelation}
\end{equation}  
which we will use to help classify the boson-impurity system.  It
should be noted that Eq.\ (\ref{eq:universalFS2}) has been defined by
others \cite{kwok08, ning08} with the exponent of $N$ being $\nu$
instead of $1/\nu$ which we have here. In the next section we numerically evaluate the FS and guided by the above scaling hypotheses find the critical exponents by collapsing the data onto universal curves.

\begin{figure}[t]
\begin{center}
\includegraphics[width=0.8\columnwidth]{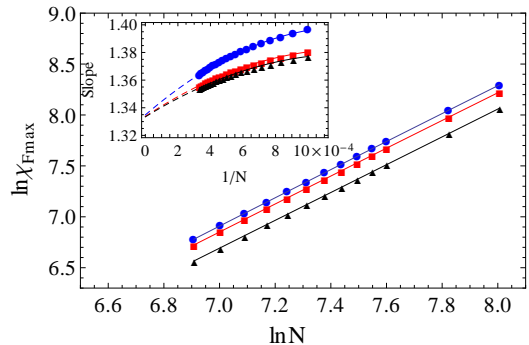}
\end{center}
\caption{(Color online) A log-log plot of $\chi_{\mathrm{Fmax}}$ as a function of $N$
  for different values of $J^a$:  $0.75 \, J$ (red squares), $1 \, J$
  (blue circles), and $1.25 \, J$ (black triangles).  Inset: The slopes
  of the log-log plot as a function of $1/N$ extrapolated in the $1/N
  \rightarrow 0$ limit.  The range of system sizes is $1000 \leq N
  \leq 3000$.}
	\label{fig:muExppic}
\end{figure}

\section{Numerical Results}
\label{sec:nr}

\begin{figure}[t]
\begin{center}
\includegraphics[width=0.8\columnwidth]{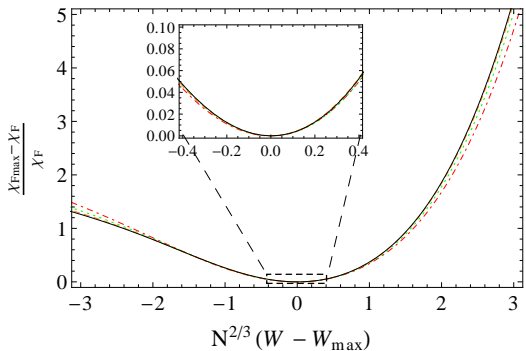}
\end{center}
\caption{(Color online) A plot of Eq.\ (\ref{eq:universalFS2}) for different system
  sizes.  The parameters and values of $N$ are the same as those used in
  Fig.\ \ref{fig:FS2}.  A value of $\nu \simeq 3/2$ results in optimal
overlay of the curves.  The inset shows magnification of the region
around the origin.}
	\label{fig:nuplot}
\end{figure} 

Our results in this section are obtained by numerically diagonalizing
the Hamiltonian given in Eq.\ (\ref{eq:ham}).  An $N$ boson system
produces a $(2N+2) \times (2N+2)$ matrix, so a system size of $N \sim
1000$ can be easily accommodated allowing us to obtain exact results.
We note that due to symmetry parity is a conserved quantity, i.e. $[\hat{H},\hat{P}] = 0$, and
hence all the eigenvectors of our Hamiltonian are either even or odd in Fock space.
Since we perform FS calculations on the ground state
(which is of even parity) we can reduce the computation time by only
considering even parity states.  However, above $W_{c}$ the eigenstates typically come in even and odd pairs separated by an exponentially small energy difference and numerical diagonalization routines find it very hard to identify the parity of such eigenvectors. Unless one is careful numerical errors lead to eigenvectors with broken symmetry \cite{mulansky11}, and this directly impacts our results since it is the critical region we are concerned with in our
calculations.  We have outlined the resolution to this problem in the
Appendix of our previous work \cite{mumford14} where we force the
eigenstates to have definite parity by diagonalizing the Hamiltonian
in the parity basis.

Figure \ref{fig:FS2} shows the results of plugging the numerically
calculated eigenstates and energies for different system sizes into
Eq.\ (\ref{eq:FS}).  We observe
a clear peak in the FS for each value of $N$ which increases in height
and sharpness as $N$ increases.  This corresponds to the shrinking of
the critical region and $W_{\mathrm{max}} \rightarrow W_c$ as $N
\rightarrow \infty$.  To find $\mu$ we first make a log-log plot of $\chi_{\mathrm{Fmax}}$ as a
function of $N$ as shown in Fig.\ \ref{fig:muExppic}.  We fit the
curves to a second degree polynomial and extrapolate their slopes in the limit $1/N \rightarrow 0$.
From the inset we see that the slopes converge to a value of $\mu
\simeq 4/3$.  We calculate $\mu$ for different values of $J^a$ to show
that $\mu$ does not depend on $J^a$ and therefore is universal.  Next, we use Eq.\ (\ref{eq:universalFS2}) to find $\nu$ by changing it
in small increments until the average overlay of data points for
different values of $N$ is maximized.  Figure \ref{fig:nuplot} shows the scaled $\chi_{\mathrm{F}}$ in the vicinity of
$W_{\mathrm{max}}$ where a maximum overlay is achieved for $\nu
\simeq 3/2$.   Figure \ref{fig:FS2} shows that below
$W_{\mathrm{max}}$ $\chi_{\mathrm{F}}$ is an intensive quantity, so we
have $d = 0$ in Eq.\ (\ref{eq:max}).  Above $W_{\mathrm{max}}$
$\chi_{\mathrm{F}}$ has a linear dependence on $N$, so
$\chi_{\mathrm{F}}/N$ is an intensive quantity and $d = 1$.  
Using Eq.\ (\ref{eq:scalerelation}) to calculate $\alpha_{\pm}$ we obtain
$\alpha_{-} \simeq 2$ and $\alpha_{+} \simeq 1/2$. These values of $\alpha_{\pm}$, $\mu$ and $\nu$ (keeping in mind the different definitions of $\nu$) are the same as those obtained
for the Lipkin-Meshkov-Glick (LMG) model numerically \cite{kwok08} and
analytically \cite{dusuel04}, for the Dicke model obtained numerically
\cite{liu09}, as well as for the system consisting of bosons in a
double well potential with attractive interactions obtained analytically \cite{buonsante12}.  This suggests that the boson-impurity system belongs in the same universality class as these models and that the QPT is second order.    

\begin{figure}[t]
\begin{center}
\includegraphics[width=0.7\columnwidth]{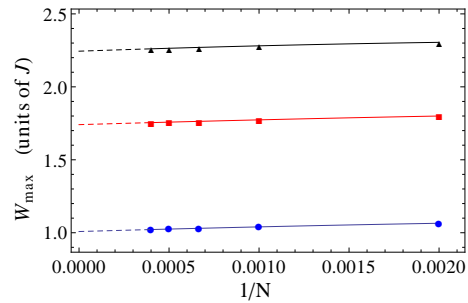}
\end{center}
\caption{(Color online) Extrapolated values of $W_{\mathrm{max}}$ for different values
  of $J^a$:  $0.75 \, J$ (red squares), $1 \, J$ (blue circles),
  and $1.25 \, J$ (black triangles).  The dashed lines show quadratic
  fits for $1/N \rightarrow 0$ and the system size range is $500 \leq
  N \leq 2500$.}
	\label{fig:Extrapo}
\end{figure}

We now shift our focus back to the convergence of $W_{\mathrm{max}}$ to
$W_c$ in the thermodynamic limit.  Using the same steps used to
determine $\mu$ we find the slope of a log-log plot of $\vert W_c -
W_{\mathrm{max}} \vert^{\delta}$ as a function of $N$ giving the
convergence scaling exponent, $\delta$, which we find to be the same as the
inverse of the
correlation length exponent, so $\delta = 1/\nu \simeq 2/3$.  In Fig.\
\ref{fig:Extrapo} we show the effectiveness of the FS in predicting $W_c$
with $1/N$ extrapolation.  For three different values of $J^a$, using
Eq.\ (\ref{eq:crit}), we
have $W_c = 1, \sqrt{3}, \sqrt{5}$ compared to the extrapolated
values of $W_{\mathrm{max}} = 1.0062, 1.7387,
2.2432$ (all values are in units of $J$).  With only five points of
data we find the two sets of values to be in good agreement.  Thus, if
we were unable to find $W_c$ analytically, the FS would provide
an excellent avenue to determine values numerically.  We summarize our
numerical results in Table \ref{tab:SE} where the uncertainties are
standard errors using a least-squares fit to our data.

\begin{center}
\begin{table*}
\begin{tabular}{ c c c c }
  \hline   
  \hline                     
  $J^a$ & 0.25 (circle) & 0.75 (square) & 1.25 (triangle) \\
  \hline
  $\mu$ & 1.335(3) & 1.334(2) & 1.333(2) \\
  $\nu$ & 1.499(2) & 1.504(5) & 1.502(3) \\
  \hline  
  $W_c$ & 1 & $\sqrt{3}$ & $\sqrt{5}$ \\
  $W_{\mathrm{Extrap}}$ & 1.0062(2) & 1.7387(3) & 2.2432(3) \\
  \hline
  \hline
\end{tabular}
\caption{Critical scaling exponents and analytic and extrapolated
  values of the QCP for different values of $J^a$.  The scaling
  exponents and QCP values are calculated with system size ranges of
  $1000 \leq N \leq 3000$ and $500 \leq N \leq 2500$, respectively.}
\label{tab:SE}
\end{table*}
\end{center}

\section{Analytic calculation of $\alpha_{\pm}$}
\label{sec:ar}

In the thermodynamic limit the critical region collapses to a point
and fluctuations vanish away from this point.  For large systems away
from the critical region this property allows us to use a mean-field approximation to
analyze the FS.  In previous work \cite{mulansky11} we have shown
that the mean-field Hamiltonian corresponding to Eq.\ (\ref{eq:ham}) is 
\begin{equation}
\frac{H_{\mathrm{MF}}}{N} =  -J \sqrt{1-Z^2} \, \mathrm{cos} \beta -
  J^a \sqrt{1-Y^2} \, \mathrm{cos} \alpha + \frac{W}{2} Z Y \ .
\label{eq:mfham}
\end{equation}
In $H_{\mathrm{MF}}$
 we have defined $\beta \equiv \beta_R - \beta_L$ and $Z \equiv \Delta N/N$ as
the boson phase and number difference between the two wells, respectively,
and $\alpha \equiv \alpha_R - \alpha_L$ and $Y \equiv \Delta M$ are
similarly defined for the impurity. The conjugate nature of the number and phase variables 
means that Hamilton's equations take the form
\begin{eqnarray}
\dot{\alpha} = \frac{1}{\hbar} \frac{\partial H}{\partial Y}  \quad ; \quad \dot{Y} =  - \frac{1}{\hbar} \frac{\partial H}{\partial \alpha} \\
\dot{\beta} = \frac{1}{\hbar} \frac{\partial H}{\partial Z}  \quad ; \quad  \dot{Z} =  - \frac{1}{\hbar}\frac{\partial H}{\partial \beta}
\end{eqnarray}
and the stable stationary solutions (which includes the ground state) of the system are 
\begin{equation}
  \left ( \alpha, Y, \beta, Z \right ) = \begin{cases}
    \left (0,0,0,0 \right ) & , \  W  \leq W_c \\
     \left ( 0,  \pm \frac{1}{W}
  \sqrt{\frac{W^4-16J^2J^{a^2}}{W^2+4J^{a^2}}}, \right. \\
\hspace{0.2cm} \left.0, \mp \frac{1}{W}
  \sqrt{\frac{W^4-16J^2J^{a^2}}{W^2+4J^2}} \right) & , \   W
 > W_c \, .
\end{cases}
\label{eq:sols}
\end{equation}
Note that for simplicity we have only displayed the solutions for the case when $W>0$ corresponding to a repulsive boson-impurity interaction.
An intuitive understanding into the role of the impurity can be gained if we use the solutions in Eq.\ (\ref{eq:sols}) to simplify Eq.\ (\ref{eq:mfham})
by adiabatically eliminating the impurity with the relation
\begin{equation}
Y = -Z \sqrt{\frac{W^2+4J^2}{W^2+4J^{a^2}}} \,
\label{eq:Ysol} 
\end{equation}
giving us an effective Hamiltonian for the bosons alone
\begin{equation}
\frac{H_{\mathrm{eff}}}{N} = -J \sqrt{1-Z^2} - J^a \sqrt{1 - Z^2 \gamma^2} -
\frac{W \gamma}{2} Z^2 \, ,
\label{eq:eff}
\end{equation}
where $\gamma = \sqrt{\frac{W^2+4J^2}{W^2+4J^{a^2}}}$.  Setting $J^a =
J$ for further simplification and scaling Eq.\ (\ref{eq:eff}) by $2 J$ gives an 
effective Hamiltonian dependent on a single parameter, $\Sigma = W/J$,
\begin{equation}
\frac{H_{\mathrm{eff}}}{2 N J} = - \frac{\vert \Sigma \vert}{4} Z^2 - \sqrt{1 - Z^2}
\, .
\label{eq:scaledeff} 
\end{equation}
A mean-field Hamiltonian of the same form occurs in the case of a
purely bosonic Josephson junction where the microscopic origin of $\Sigma$ is
direct boson-boson interactions \cite{milburn97,smerzi97}. Specifically, the minus sign in front of the first term indicates effectively attractive boson-boson interactions.  Although we have calculated $H_{\mathrm{eff}}$ 
here assuming repulsive boson-impurity interactions, it turns out to be unchanged for
attractive interactions.  Thus, the impurity always mediates attractive effective boson-boson interactions \cite{heiselberg00,santamore08}, and it is for this reason that the PT in the impurity model falls into the same universality class as the clumping PT for attractive bosons. We can visualize
how this happens by considering the
impurity localized in one well and having $\vert W \vert > W_c$, so the
ground state will have a larger fraction of bosons in one well over
the other.  For $W > 0$ the impurity expels bosons from the well it's
in and for $W < 0$ bosons are attracted to the impurity.  In both
cases there is a build-up of bosons in one well over the other which
is what happens when there are attractive boson-boson interactions.

An analytic calculation of the scaling exponents for the clumping transition for attractive bosons has been given in reference \cite{buonsante12}. Their method for calculating the FS consists of approximating the ground state
wave function as a Gaussian in Fock space centered at $Z = 0$ for $W \ll W_c$ and a
symmetric superposition of Gaussians for $W \gg W_c$.  In our
calculations we do not use a superposition of Gaussians for $W \gg
W_c$, but instead choose to have a single Gaussian centered at one of
the two mean-field solutions, shown in Eq.\ (\ref{eq:scaledmin}), to
represent the symmetry broken phase.  The difference in these two
approaches results in terms proportional to $e^{-N \vert \Sigma -
  \Sigma_c \vert}$, so if we are sufficiently far from the critical
region, then each approach is equivalent.  Using a
different form of the FS \cite{cozzini,buonsante12}
\begin{equation}
\chi_{\mathrm{F}}(\Sigma) = -\frac{1}{2} \frac{d^2}{d \delta \Sigma^2} \langle
\psi_0(\Sigma) \vert \psi_0(\Sigma + \delta \Sigma) \rangle
\mid_{\delta \Sigma =0}
\label{eq:newfid}
\end{equation}
they are able to calculate analytic expressions for the FS.  Following
their steps for Eq.\ (\ref{eq:scaledeff}), which we briefly outline in
Appendix \ref{sec:analsteps},
we obtain

\begin{equation}
\chi_{\mathrm{F}}(\Sigma) = \begin{cases}
  \frac{1}{64 \left ( \Sigma - 2 \right )^2} & , \  \Sigma \ll \Sigma_c \\
  \frac{N}{\vert \Sigma \vert^3 \sqrt{2 (\Sigma^2 - 4)}} + \frac{
    \left (\Sigma^2 - 2
    \right )^2}{4 \Sigma^2 \left ( \Sigma^2 - 4 \right )^2} & , \ 
  \Sigma \gg \Sigma_c \, .
\end{cases}
\label{eq:analchi}
\end{equation}
We can see the scaling exponents are $\alpha_- = 2$ and $\alpha_+ =
1/2$ agreeing with the numerical values calculated in the
previous section.  Equation (\ref{eq:analchi}) shows the leading order
behaviour of the FS.  Below $\Sigma_c$ there is a single leading term
because the Gaussian wave function is fixed at $Z = 0$, so changes in
$\Sigma$ can only affect its size.  Above $\Sigma_c$ changes in
$\Sigma$ affect both the size and position of the wave function giving
two terms where we see in the thermodynamic limit the position
dependent term dominates.

\section{Summary and Discussion}
\label{sec:dc}

In this paper we have studied a symmetry breaking bifurcation in a bosonic Josephson junction driven by the interaction with an impurity atom. The fact that the maximum value of the FS, which can be viewed as a generalized susceptibility, diverges in the thermodynamic limit confirms that the symmetry breaking is associated with a second order phase transition (as expected from the continuous form of the bifurcation). By numerically calculating the critical scaling exponents of the FS and comparing them with those already known in the Dicke and LMG models, as well as for a system consisting of bosons in a double well potential with attractive interactions, we conclude that the PT in the impurity model lies in the same universality class as these other models. For the two exponents $\alpha_{\pm}$ of the scaling of FS with $W$ on either side of the transition, we also carried out an analytic calculation and good agreement was found with the numerical result. We have also shown through extrapolation of $W_{\mathrm{max}}$ in the thermodynamic limit that the FS can be used to predict $W_c$ numerically, and we find that it agrees with the analytic result calculated from the mean-field theory.

Interpreting the bosons as a meter measuring the position of the impurity, we have a particularly simple toy model for a binary quantum measurement in terms of a PT which occurs at a critical value of the system-meter interaction strength \cite{damnjanovic,mayburov,allahverdyan,bargill,ivanov}.  Quantum mechanically, the ground state probability distribution goes from having Gaussian fluctuations around $\Delta N = 0$ to a superposition of two Gaussians each centered at one of the two bifurcating mean-field solutions.  This latter state becomes a Schr\"{o}dinger Cat state if $N \gg 1$ and $W > W_c$. Cat states are notoriously sensitive to perturbations and can be expected to rapidly collapse into one of the two wells thereby breaking the symmetry. This collapse is implicit in our model but it is interesting to ask whether a third agent beyond the impurity and the bosons is necessary to precipitate it. If the symmetry is broken by a classical field then it can be simply included in the Hamiltonian as a tilt to the double well potential \cite{rinck11,mulansky11,mumford14} and as long as the perturbation is infinitesimal the PT is not affected. However, if the boson-impurity system is instead put into contact with a quantum mechanical environment then the effects can be more marked. PTs in open quantum systems (systems coupled to an environment) are now the subject of intensive research \cite{opencrit,goldbart09}, especially for the open Dicke model  \cite{nagy10,bhaseen12,dickecrit0,sachdev}. One conclusion of this body of work is that the critical exponents can be modified by the coupling to the environment and this effect has been seen experimentally \cite{dickecrit2}.

Finally, we mention that the impurity localization described in this paper is somewhat different to that found in the classic problem of an impurity in a uniform superfluid \cite{miller}, or its modern descendant, an impurity in an extended gaseous Bose-Einstein condensate (BEC) \cite{kalas06,cucchietti06,tempere09,cirone09,compagno14}. For example, the Bose-Hubbard Hamiltonian employed here is a tight-binding model where the single particle wave functions (modes) are assumed to be unchanged by interactions, whereas the transition to a self-localized polaron state in an initially uniform BEC involves a change in the impurity wave function from delocalized to localized and the BEC develops a corresponding density dip.  Furthermore, the type of symmetry that is broken in going from a uniform to a localized wave function is in general different to the binary choice underlying $\mathbb{Z}_{2}$ symmetry breaking (see Reference \cite{dhar14} for the case of a particle living on one and two-dimensional lattices with many lattice sites). However, in one dimensional extended systems the Josephson model underlying the physics studied here appears quite naturally as the impurity splits the BEC in two and we would expect there to be connections \cite{dima1,dima2}. We also point out that there are many other aspects to the impurity model and its close relatives beyond those discussed here, including how the coherence of the bosons is affected by the impurity \cite{bausmerth,mulansky11,lu12}, and system-bath dynamics \cite{spehner,ferrini,mcendoo,kronke14}.

\begin{acknowledgments}
We acknowledge insightful discussions with Jonas Larson and Sung-Sik Lee.  This research was funded by the Natural Sciences and Engineering Research Council (NSERC) of Canada. JM also acknowledges funding from the government of Ontario.
\end{acknowledgments}

\appendix
\section{Steps for Analytic Calculations}
\label{sec:analsteps}

In this appendix we briefly outline the steps used to derive Eq.\
(\ref{eq:analchi}) from Eq.\ (\ref{eq:scaledeff}).  We start by
expanding Eq. (\ref{eq:scaledeff}) around the minima above and below
$\Sigma_c$
\begin{equation}
Z_0 = \begin{cases}
0 & , \ \Sigma \leq \Sigma_c \\
\pm \sqrt{1 - \left ( \frac{2}{\Sigma}  \right )^2} & , \  \Sigma
> \Sigma_c
\end{cases}
\label{eq:scaledmin}
\end{equation}
where $\Sigma_c = 2$.  If we are sufficiently far away from
$\Sigma_c$, then $H_{\mathrm{eff}}$ is parabolic in shape around the
minima, so the leading order term in the expansion will be the
second giving a Schr\"{o}dinger equation  
\begin{equation}
\left [ -\frac{d^2}{du^2} + h(\Sigma) u^2 \right ] \Psi_\Sigma(Z) = E \Psi_\Sigma(Z)
\label{eq:schro}
\end{equation}
where $u = Z - Z_0$ and 
\begin{equation}
h(\Sigma) = \begin{cases}
\frac{N^2}{4} \left (- \Sigma +2  \right ) & , \ \Sigma \ll \Sigma_c \\
\frac{N^2}{32} \Sigma^2 \left ( \Sigma^2 - 4  \right ) & , \ \Sigma \gg
\Sigma_c
\end{cases} \, .
\label{eq:h}
\end{equation}
Equation (\ref{eq:schro}) describes a harmonic oscillator in
Fock space which means the ground state wave function will be a
Gaussian of the form
\begin{equation}
\Psi_{\Sigma}(Z) = \frac{1}{\sqrt{\sigma_\Sigma \sqrt{2
      \pi}}} e^{-\frac{(Z-Z_0)^2}{4 \sigma^2_{\Sigma}}}
\label{eq:gssmallsig} \, .
\end{equation} 
The difference between the $\Sigma < \Sigma_c$ and $\Sigma > \Sigma_c$
wavefunctions is due to $Z_0$ through Eq.\ (\ref{eq:scaledmin}) and the relation
$\sigma^2_{\Sigma} = \frac{1}{2 \sqrt{h(\Sigma)}}$.  With these forms
of the ground state we can use Eq.\ (\ref{eq:newfid}) giving 
\begin{equation}
\chi_F(\Sigma) = -\frac{1}{2} \frac{d^2}{d \delta \Sigma^2}
\int_{-\infty}^{\infty} \Psi_{\Sigma}(Z) \Psi_{\Sigma + \delta
  \Sigma}(Z) dZ \mid_{\delta \Sigma = 0} \, 
\end{equation}
and from here we obtain the expressions given in Eq.\ (\ref{eq:analchi}).


\begin{thebibliography}{8}
\bibitem{note1}{In other words, when $1/(E_{1}-E_{0})$ is evaluated at $W=W_{\mathrm{max}}$ using the eigenvalues of Eq.\ (\ref{eq:ham}) as it stands it scales as $N^{1/3}$ as we have numerically verified.}
\end{thebibliography}
\end{document}